\documentclass{article}

\date{}

\usepackage{hanmin,amsmath,amssymb}
\newtheorem{theorem}{Theorem}
\newcommand{\D}{\mathrm{D}}
\newcommand{\Epsilon}{\mathcal{E}}

\begin{document}

\title{A simple proof of a duality theorem with applications in scalar and anisotropic viscoelasticity}

\author{Andrzej Hanyga\\
ul. Bitwy Warszawskiej 1920 r 14/52\\
02-366 Warszawa, Poland\\
ajhbergen@yahoo.com}


\begin{abstract}
A new concise proof is given of a duality theorem connecting completely monotone relaxation functions 
with Bernstein class creep functions in one-dimensional
and anisotropic 3D viscoelasticity. The proof makes use of the theory of complete Bernstein functions and 
Stieltjes functions and is based on a relation between these two function classes. 
\end{abstract}

\maketitle

\section*{Notations}
$[0,\infty[ := \{ x \in \mathbf{R} \mid 0 \leq x < \infty \}$\\
$]0,\infty[ := \{ x \in \mathbf{R} \mid 0 < x < \infty \}$\\ 
$\D f(t) := \dd f(t)/\dd t = \dot{f}(t)$\\
\vspace{0.2cm}

\section{Introduction.}

It happens so that all the mathematical models of viscoelastic relaxation moduli are completely
monotone (CM) functions. This is partly due to the interpolation procedures applied in
deriving the relaxation moduli (such as Prony sums), or, more often, the creep functions from discrete experimental data.
The set of CM functions is a very thin subset in the spaces of continuous
or smooth fuctions, i. e. in an arbitrarily small neighborhood of a CM function in the $\mathcal{C}^0$ 
topology there are functions which are not CM, but the interpolation procedures are based a priori on CM functions
such as the exponential function $\exp(-k t)$.
By the duality theorem considered below the corresponding creep functions
are Bernstein functions (BFs).  

The assumption of the CM property of relaxation moduli has very powerful consequences especially for
viscoelastic wave propagation \cite{HanWM,HanAnisoWaves}. Therefore it is convenient to stick to this assumption so long as it does
not contradict experimental data.

In this paper I shall 
demonstrate the application of the complete Bernstein functions and the Stieltjes functions 
in the proof of a well known duality relation between locally integrable completely monotone (LICM) 
relaxation moduli $R$ and Bernstein class creep functions $C$. I recall the viscoelastic
duality relation \cite{Molinari73,HanISIMM,MainardiBook}
\begin{equation} \label{duality}
\tilde{R}(p) \, \tilde{C}(p) = p^{-2}
\end{equation}
where 
\begin{equation}\label{Laplace}
\tilde{f}(p) := \int_0^\infty \e^{-p t} \, f(t) \, \dd t
\end{equation}
denotes the Laplace transform
for any locally integrable function $f$ such that the transform exists.

Experimental evidence suggests that the relaxation moduli of all the known viscoelastic media are completely monotone
functions of time.
It was previously shown by other methods \cite{Molinari73,HanISIMM,HanDuality} that, apart from some singular terms, 
if $R$ is locally integrable completely monotone (LICM),  then the creep function $C$ is a Bernstein function and conversely.
The new proof of the same relation is more elegant and concise than the 
previously given proofs \cite{Molinari73,HanISIMM,HanDuality} of the same relation. It is based on the theory of 
complete Bernstein functions and Stieltjes functions \cite{SchillingAl}. 

The main advantage of this approach is availability of integral representations of Stieltjes and complete Bernstein functions.
As soon as we have established that a function of interest is a Stieltjes or complete Bernstein function
we have a decomposition of that function in the form of the sum of three terms which is complete. For example we discover
that for a given creep function the dual relaxation function must
contain a Newtonian viscosity term.

In \cite{HanDuality} the results of \cite{HanISIMM} were extended to tensor-valued relaxation moduli and creep functions.
In this paper we shall also study the duality relation for tensor-valued relaxation moduli and creep functions.
The tensor-valued functions will be treated as matrix-valued functions on $\mathbb{R}^6$. We shall use the theory 
of matrix-valued complete Bernstein and Stieltjes functions developed in \cite{HanAnisoWaves}.

While in the scalar case we obtain a fairly complete set of relations between relaxation and creep, 
in the anisotropic case there are some limitations due to the complex questions of matrix invertibility.

These relations are of paramount importance for deducing the relaxation modulus from creep test data.

\section{Preliminaries.}

It is convenient for our considerations to consider functions $f$ in 
$\mathcal{L}^1_{\mathrm{loc}}([0,\infty[)$ as convolution operators $f\ast$ mapping 
$g \in \mathcal{L}^1_{\mathrm{loc}}([0,\infty[)$ to $f\ast g \in \mathcal{L}^1_{\mathrm{loc}}[0,\infty[)$.

The convolution of two locally integrable functions $f$ and $g$ on $[0,\infty[$
is defined by the formula
\begin{equation}
(f\ast g)(t) := \int_0^t f(s) \, g(t-s) \, \dd s
\end{equation}
If the Laplace transforms $\tilde{f}(p)$ and $\tilde{g}(p)$ exist for some $p \geq 0$, then the convolution 
$f\ast g$ also has the Laplace transform at $p$ and 
\begin{equation} \label{LaplId}
(f\ast g)\,\widetilde{}\,(p) = \tilde{f}(p)\, \tilde{g}(p) 
\end{equation}

We shall also need an identity operator on $\mathcal{L}^1_{\mathrm{loc}}([0,\infty[)$:
\begin{equation} \label{unity}
\mathrm{U}\, f = f
\end{equation}
For the sake of convenience we shall also write \eqref{unity} in the form
\begin{equation}
u\ast f = f
\end{equation}

Extending \eqref{LaplId} to \eqref{unity} we have
$\tilde{u}(p) \tilde{f}(p) = \tilde{f}(p)$, whence
\begin{equation}
\tilde{u}(p) = 1,\; p \geq 0
\end{equation}

Note that if $\sigma = R\ast \dot{\epsilon}$, where $\sigma$ represents the stress, 
$\dot{\epsilon}$ is the strain rate and the relaxation modulus $R = N \, u + f_0$, where 
$f_0 \in \mathcal{L}^1_{\mathrm{loc}}([0,\infty[)$, then $\sigma = N \, \dot{\epsilon} + 
f_0\ast \dot{\epsilon}$. In the absence of the second term the first term  represents Newtonian viscosity.
We shall however see that for the validity of the duality relation the appearance of a term $b \, u $ is necessary.

The mathematical function classes required in the proof of the duality theorem are 
explained in the appendix along with their properties which will be needed in the
following.

\section{Scalar viscoelasticity.}

\begin{theorem}
If $f_0$ is LICM and not identically zero and 
\begin{equation} \label{f0}
f(t) = f_0(t) + \beta \, u(t)
\end{equation}
where $\beta \geq 0$, then 
\begin{equation} \label{y2} 
1/[p \tilde{f}(p)] = p\, \tilde{h}(p)
\end{equation} 
where $h$ is a Bernstein function.

We also have $0 < f_0(0) \leq \infty$. If $\beta > 0$ or $f_0(t)$ is unbounded at 0 then
$h(0) = 0$, otherwise $h(0) = 1/f_0(0)$.

The limit $f_\infty : = \lim_{t\rightarrow \infty} f_0(t)$ exists and is non-negative. \\
If $f_\infty > 0$, then for $t \rightarrow \infty$ the function $h(t)$ tends to $1/f_\infty$, 
otherwise it diverges to infinity.
\end{theorem}

\noindent\textbf{Proof.} 

We shall use a few theorems on  of the CBFs and Stieltjes functions quoted in the appendix to construct the Bernstein 
function $h$.

Since $f_0$ is LICM, there is a non-negative real number $a$ and Borel measure $\mu$ on $]0,\infty[$ satisfying the inequality
\begin{equation} \label{x2}
\int_{]0,\infty[} (1 + s)^{-1} \,\mu(\dd s) < \infty
\end{equation}
such that 
\begin{equation} \label{vv}
f_0(t) = a + \int_{]0,\infty[} \e^{-s t} \mu(\dd s)
\end{equation}

It follows that
\begin{equation}
\tilde{f}_0(p) = a/p + \int_{]0,\infty[} (s + p)^{-1} \, \mu(\dd s) 
\end{equation}
hence, by eq.~\eqref{app2}, $p\, \tilde{f}(p) = \beta\, p + p\, \tilde{f}_0(p)$ is a 
CBF and $1/[p \tilde{f}(p)]$  is a Stieltjes function.
Hence there are non-negative real numbers $a, b$ and a Borel measure $\nu$ on $]0,\infty[$ satisfying the inequality
\begin{equation} \label{x0}
\int_{]0,\infty[} (1 + s)^{-1}\,\nu(\dd s) < \infty
\end{equation}
such that
\begin{equation}\label{x11}
1/[p \, \tilde{f}(p)] = a + \frac{b}{p} + \int_{]0,\infty[} \frac{\nu(\dd r)}{r + p} 
\end{equation}

The last term is the Laplace transform $\tilde{g}(p)$ of the LICM function
\begin{equation}
g(t) := \int_{]0,\infty[} \e^{-t r} \, \nu(\dd r)
\end{equation}

If
$$G(t) := \int_0^t g(s)\, \dd s + a$$
then
$$p \, \tilde{G}(p) = \tilde{g}(p) + a$$
and $G$ is a Bernstein function. 

Concerning the term $b/p$ appearing in \eqref{x11} we note that  
$b/p^2$ is the Laplace transform of the Bernstein function $b \, t, t \geq 0$. 
Hence eq.~\eqref{y2} holds with the Bernstein function 
\begin{equation} \label{ht}
h(t) := b\, t + G(t) 
\end{equation}

It remains to consider the limits of these functions.

The function $f_0(t) \geq 0$ is non-increasing and $f_0 \not\equiv 0$.  
$f_0(t)$ may be unbounded at 0, otherwise the limit $f_0(0)$ of $f_0(t)$ at 0 exists and $f_0(0) > 0$. On the other hand $h(0) \geq 0$
is always finite.
On account of equation~\eqref{y2} 
$$h(0)  = \lim_{p\rightarrow \infty} 1/[\beta \, p + p \, \widetilde{f_0}(p)] $$ 
If $\beta > 0$ or $f_0(t)$ is unbounded at 0, then $h(0) = 0$, otherwise $h(0) = 1/f_0(0)$.

The function $f_0$ satisfies the inequalities $0 \leq f_0(t) \leq f_0(1)$ for $t > 1$ and its non-increasing. 
Consequently the limit $f_\infty := \lim_{t\rightarrow \infty} f_0(t)$ exists and is non-negative. We now note that 
$$\lim_{p\rightarrow 0} [p \, \tilde{f}(p)] = \lim_{p\rightarrow 0} [p \, \widetilde{f_0}(p))] = f_\infty,$$
If $f_\infty > 0$ then 
$$\lim_{t\rightarrow \infty} h(t) = \lim_{p \rightarrow 0} [p \, \tilde{h}(p)] = 1/f_\infty$$
on account of \eqref{y2}. Otherwise $h(t)$ is unbounded at infinity and 
the limit $\lim_{t\rightarrow \infty} h(t)$ does not exist.
\\
$\mbox{ }$ \hfill $\Box$\\

\begin{theorem}
If $h$ is a B not identically zero, then there is a LICM function $f_0$ and a real 
$b \geq 0$ such that 
\begin{equation} \label{y4}
1/[p\, \tilde{h}(p)] = p \, \tilde{f}(p)
\end{equation}
where
\begin{equation}
f(t) := b \, u(t) + f_0(t)
\end{equation}

The function $h(t)$ is either bounded and tends to a positive limit at infinity or it diverges to infinity.
In the first case we have the identity
$$\lim_{t\rightarrow \infty} f_0(t) = \lim_{t \rightarrow \infty} 1/h(t),$$
otherwise $f_0(t)$ tends to 0 at infinity.

If $h(0) > 0$, then $f_0(t)$ is bounded at 0,  $f_0(0) = 1/h(0)$ and $b = 0$,
otherwise either $b > 0$ or $f_0(t)$ is unbounded at 0. 

If $h(0) = 0$, then $b = 1/h^\prime(0)$ for $h^\prime(0) \geq 0$.

\end{theorem}

\noindent\textbf{Proof.}

Since $p \, \tilde{h}(p) = [h^\prime]\tilde(p) + h(0)$ and the derivative $h^\prime$ \
of $h$ is LICM, $p \, \tilde{h}(p)$ is a Stieltjes function and therefore 
$1/[p \, \tilde{h}(p)]$ is a CBF.

Eq.~\eqref{app2} implies that there are two reals $a, b \geq 0$ and a Borel measure $\nu$ satisfying 
\eqref{x0} such that
$$1/[p \, \tilde{h}(p)] = a + b\, p + p \int_{]0,\infty[} \frac{\nu(\dd r)}{r + p}$$
Let 
$$f_0(t) := a + \int_{]0,\infty[} \e^{-r t}\, \nu(\dd r)$$
and
$$f(t) := f_0(t) + b \, u(t)$$
$f_0$ is clearly LICM and eq.~\eqref{y4} is satisfied.

Furthermore, $\lim_{t\rightarrow \infty} h(t) > 0$ and 
\begin{equation}
\lim_{t \rightarrow \infty}\, 1/h(t) = \lim_{p \rightarrow 0} 1/[p \, \tilde{h}(p)]
= \lim_{p\rightarrow 0} [p \, \tilde{f}(p)] = \lim{p\rightarrow 0} \widetilde{f_0}(p) = \lim_{t\rightarrow \infty}\, f_0(t)
\end{equation}

At the other end we note that $h(0) \geq 0$ exists. If $b = 0$, and $f_0(t)$ is bounded at 0,  then
$$h(0) = \lim_{p\rightarrow\infty} [p \, \tilde{h}(p)] = \lim_{p\rightarrow\infty} 1/[b \, p + p\, \widetilde{f_0}(p)] 
= 1/f_0(0),$$ otherwise $h(0) = 0$.

Hence if $h(0) > 0$, then $b = 0$ and $f_0(t)$ tends to $1/h(0)$ for $t \rightarrow 0$,
while if $h(0) = 0$ then either $b > 0$ or $f_0(t)$ diverges to infinity at 0.

If $h(0) = 0$, then $\lim_{p\rightarrow \infty} [p^2 \, \tilde{h}(p)] = \lim_{p\rightarrow \infty} [p \, \widetilde{h^\prime}(p)]
 = \lim_{t\rightarrow 0} h^\prime(t)$. The last limit exists because $h^\prime$ is LICM, but it may be infinite. 
On the other hand \eqref{y2} implies that $\lim_{p\rightarrow \infty} [p^2 \, \tilde{h}(p)] =
\lim_{p\rightarrow \infty} 1/[b + \widetilde{f_0}(p)]$. We now note that
\begin{equation} \label{y1} 
\lim_{p\rightarrow \infty} \widetilde{f_0}(p) = \lim_{p\rightarrow \infty} \{p \, [1/p \widetilde{f_0}(p)]\} =
\lim_{t\rightarrow 0} \int_0^t f_0(t) \, \dd t = 0
\end{equation}

The last equation in \eqref{y1} follows from the fact that $f_0$ is integrable over $[0,1]$.
Indeed,  for $t \leq 1$  
$$\int_0^t f_0(t) \, \dd t = \int_0^1 f_0(s) \, \theta(t - s) \, \dd s$$
where $\theta$ denotes the unit step function, $f_0(s)  \, \theta(t - s) \leq f_0(s)$ and 
$f_0(s)  \, \theta(t - s) \rightarrow 0$ for $0 \leq s \leq 1$ as $t \rightarrow 0$,
hence the last equation in \eqref{y1} follows from the Lebesgue dominated convergence theorem.

We thus conclude that in the case 
of $h(0) = 0$ 
$$b = 1/\lim_{t\rightarrow 0} h^\prime(t).$$
with $b = 0$ if $h^\prime(t) \rightarrow \infty$ for $t \rightarrow 0$.
\\
$\mbox{ }$ \hfill $\Box$ \\

\section{Extension to anisotropic media.}

Constitutive equations of anisotropic viscoelastic media in three-dimensional space assume the following form
\begin{eqnarray}
\Sigma_{I} = \sum_{J=1}^6 R_{IJ} \ast \dot{\Epsilon_J},  \          I = 1,\ldots 6\\
\Epsilon_I = \sum_{J=1}^6 C_{IJ} \ast \dot{\Sigma_J},    \          I = 1,\ldots 6
\end{eqnarray}
where $I = 1, 2, 3$ stands for 11, 22 and 33, respectively,  and $I = 4, 5, 6$ for $23, 31, 12$, 
$\Epsilon_k = \epsilon_{kk}$ for $k=1,2,3$, $\Epsilon_{3+l} = 2^{1/2} \,\epsilon_{mn}$ for $l \neq m,n,$ and $m\neq n$,  with similar rules for $R_{IJ}$. While 
$R_{ijkl}(t)$, $C_{ijkl}(t)$ and $\epsilon_{ij}$ are tensor-valued functions, the corresponding $6$-dimensional objects are $6 \times 6$
and $6 \times 1$ matrices. 
  
Due to the major symmetry $R_{ijkl} = R_{klij}$, $C_{ijkl} = C_{klij}$ 
the functions $R_{IJ}(t)$, $C_{IJ}(t)$  defined on $]0,\infty[$ take values in the space $\mathcal{S}_+$ of positive semi-definite 
symmetric matrices $6\times 6$ matrices.
The $6\times 6$ relaxation modulus and creep function will be denoted by $\mathbb{R}(t)$ and $\mathbb{C}(t)$, respectively. 

We shall study the relation
\begin{equation} \label{anisorel}
p\, \tilde{\mathbf{R}}(p) = p^{-1}\, \tilde{\mathbf{C}}(p)^{-1}
\end{equation}
\cite{HanDuality}.

Matrix-valued CB and Stieltjes functions were studied in \cite{HanAnisoWaves}. Some results relevant for us are collected 
in Appendix~\ref{appAniso}. There is a close analogy between them and the results of Appendix~\ref{appscalar} used in the previous section.

Recall that $\mathbf{R}(t)$ at 0 and $\mathbf{C}(t)$ at infinity may be unbounded. If however $\mathbf{v}^\mathsf{T} \, \mathbf{R}(t) \,
\mathbf{v}$ is bounded for every $\mathbf{v} \in \mathbb{R}^6$, then it tends to a limit. By polarization we conclude that 
$\mathbb{R}(t)$ tends to a limit which we denote as the value $\mathbf{R}(0)$. Similarly, if the creep function
$\mathbf{v}^\mathsf{T}\, \mathbf{C}\, \mathbf{v}$ is bounded for each $\mathbf{v} \in \mathbb{R}^6$, then 
$\lim_{t\rightarrow \infty} \mathbf{C}(t)$ exists.

\begin{theorem} \label{thm1aniso}
If $\mathbf{R}(t) = u(t)\,\mathbf{N} + \mathbf{F}(t)$, where $\mathbf{F}$ is a $\mathcal{S}_+$-valued LICM and $\mathbf{N} 
\in \mathcal{S}_+$ and \\
$(\ast)\;$ for each non-zero vector $\mathbf{v} \in \mathbb{R}^6$ the function $R_{IJ}(t)\, v_I\, v_J$ is not identically zero, \\
then there is a $\mathcal{S}_+$-valued Bernstein function $\mathbf{C}$ such that equation~\eqref{anisorel} holds.

The limit $\lim_{t \rightarrow 0} \mathbf{C}(t) = \mathbf{C}(0)$ always exists and is 
positive semi-definite.

If $\mathbf{N} > 0$, then $\mathbf{C}(0) = 0$ and $\lim_{t\rightarrow 0} \mathbf{C}^\prime(t) = \mathbf{N}^{-1}$.

If $\mathbf{N} = 0$, the limit $\lim_{t\rightarrow 0} \mathbf{F}(t)$ exists and is invertible, then
$\mathbf{C}(0) = \left[ \lim_{t\rightarrow 0} \mathbf{F}(t) \right]^{-1}$.

The limit $\lim_{t\rightarrow \infty} \mathbf{R}(t) =: \mathbf{F}_\infty$ always exists and is positive-semi-definite. 

If $\mathbf{F}_\infty$ is invertible, then $\lim_{t\rightarrow \infty} \mathbf{C}(t)$ exists and 
\begin{equation} \label{eq3}
\lim_{t\rightarrow \infty} \mathbf{C}(t) = \left[\lim_{t\rightarrow \infty} \mathbf{R}(t) \right]^{-1}
\end{equation}
\end{theorem}

\noindent\textbf{Proof.}\\
On account of  \eqref{mlapLICM}
$$p \,\tilde{\mathbf{R}}(p) = p \, \mathbf{N} + \mathbf{B} + p \int_{]0,\infty[} (p + r)^{-1}\, \mathbf{G}(r) \, \mu(\dd r),$$
hence $p \, \tilde{\mathbf{R}}(p)$ is an  $\mathcal{S}_+$-valued CBF. 

On account of $(\ast)$ the matrix $\tilde{\mathbf{R}}(p)$ has an inverse for $p > 0$. Indeed, for every vector $\mathbf{v}$ there is a 
positive number $t_\ast(\mathbf{v})$ such that $\mathbf{v}^\mathsf{T}\, \mathbf{R}(t_\ast)\, \mathbf{v} > 0$, while $\mathbf{v}^\mathsf{T}\, \mathbf{R}(t)\, \mathbf{v} \geq 0$
for all $t > 0$. Hence $\tilde{\mathbf{R}}(p) > 0$  for all $p \geq 0$ and thus $\tilde{\mathbf{R}}(p)$ has an inverse for $p \geq 0$.

The inverse of $p\, \tilde{\mathbf{R}}(p$ is thus an $\mathcal{S}_+$-valued Stieltjes function and has the form
\begin{equation} \label{yu}
\mathbf{A} + p^{-1} \, \mathbf{D} + \int_{]0,\infty[} (p + r)^{-1}\,\mathbf{H}(r)\, \nu(\dd r),
\end{equation}
where $\mathbf{A}, \mathbf{D} \in \mathcal{S}_+$, $\nu$ is a Borel measure on $]0,\infty[$ satisfying \eqref{x0} and 
$\mathbf{H}$ is a bounded measurable $\mathcal{S}_+$-valued function 
defined $\nu$-almost everywhere on $]0,\infty[$.
On account of equation~\eqref{anisorel} $p \, \tilde{\mathbf{C}}(p)$ has the form given by equation~\eqref{yu}. It follows that 
$$\mathbf{C}(t) = \mathbf{A} + t\, \mathbf{D} + \int_0^t \mathbf{K}(s) \, \dd s, $$
where $$\mathbf{K}(t) := \int_{]0,\infty[} \e^{-r t}\, \mathbf{H}(r)\, \nu(\dd r)$$ is a LICM. It follows that $\mathbf{C}$ is
an $\mathcal{S}_+$-valued Bernstein function.

We now turn to the limits of the creep function. 

Note that the limit $\mathbf{F}_\infty := \lim_{t\rightarrow \infty} \mathbf{F}(t)$ always exists and is positive semi-definite. 
Indeed, the function $\mathbf{v}^\mathsf{T}\, \mathbf{F}(t)\, \mathbf{v}$ is a non-negative non-increasing for every $\mathbf{v} 
\in \mathbb{R}^6$, hence it tends to a limit. By the polarization argument $\mathbf{v}^\mathsf{T}\, \mathbf{F}(t) \, \mathbf{w}$
also tends to a limit for $t \rightarrow \infty$ for every $\mathbf{v}, \mathbf{w}$. Hence $\lim_{t\rightarrow \infty} \mathbf{F}(t)$ exists and is positive semi-definite.

By an analogous argument $\mathbf{C}(0)$ is defined and positive semi-definite.

Now
$$\lim_{t\rightarrow 0} \mathbf{C}(t) = \lim_{p\rightarrow \infty} [p\, \tilde{\mathbf{C}}(p)]  = \lim_{p\rightarrow \infty} 
\left[p \, \mathbf{N} + \mathbf{B} + p \int_{]0,\infty[} (r + p)^{-1}\, \mathbf{G}(r) \, \mu(\dd r)\right]^{-1}$$

If $\mathbf{N} > 0$, then the right-hand side can be recast in the form 
$$\lim_{p\rightarrow \infty} \left\{ p^{-1} \, \left[ \mathbf{N} + p^{-1}\, \mathbf{B} +
\int_{]0,\infty[} (r + p)^{-1} \, \mathbf{G}(r)\, \mu(\dd r) \right]^{-1} \right\}$$
The last term is the Laplace transform of the integral  
$\int_0^t \mathbf{F}_0(s)\, \dd s$, where $\mathbf{F}_0(t) := \int_{]0,\infty[} \e^{-r t} \, \mathbf{G}(r) \, \mu(\dd r)$ is a LICM function.
Hence the limit at $p\rightarrow \infty$ of that term is equal to $$\lim_{t\rightarrow 0} \int_0^t \mathbf{F}_0(s)\, \dd s,$$
which vanishes because $\mathbf{F}_0(t)$ is integrable over $[0,1]$. 
Consequently $$\lim_{p\rightarrow \infty} \left[\mathbf{N} + p^{-1}\, \mathbf{B} +
\int_{]0,\infty[} (r + p)^{-1} \, \mathbf{G}(r)\, \mu(\dd r)\right]  = \mathbf{N}$$ and is invertible.
Therefore 
$\lim_{p\rightarrow\infty} \left[\mathbf{N} + p^{-1}\, \mathbf{B} +
\int_{]0,\infty[} (r + p)^{-1} \, \mathbf{G}(r)\, \mu(\dd r)\right]^{-1}$ exists and equals $\mathbf{N}^{-1}$.
It follows that $\lim_{t\rightarrow 0} \mathbf{C}(t) = 0$ in this case. 

Furthermore, under the same assumption 
$$\lim_{t\rightarrow 0} \mathbf{C}^\prime(t)  = \lim_{p\rightarrow \infty} [p^2\, \tilde{\mathbf{C}}(p)] =
\lim_{p\rightarrow \infty} \left[\mathbf{N} + p^{-1}\, \mathbf{B} +
\int_{]0,\infty[} (r + p)^{-1} \, \mathbf{G}(r)\, \mu(\dd r)\right]^{-1} = \mathbf{N}^{-1}$$

If $\mathbf{N} = 0$, then
$$\lim_{t\rightarrow 0} \mathbf{C}(t) = \lim_{p\rightarrow \infty} \left[\mathbf{B} + p \, \int_{]0,\infty[} (r + p)^{-1}\, \mathbf{G}(r)
\, \mu(\dd r)\right]^{-1}$$ 
The limit of the last term at $p \rightarrow \infty$ is $\lim_{t\rightarrow 0} \mathbf{F}_0(t)$, which may be infinite.
However, if the last limit is finite and $\mathbf{B} + \lim_{t\rightarrow 0} \mathbf{F}_0(t) = \mathbf{R}(0)^{-1}$ is invertible, then
$\mathbf{C}(0) = \left[\mathbf{R}(0)\right]^{-1}$.

Finally, 
$$\lim_{t\rightarrow \infty} \mathbf{C}(t) = \lim_{p\rightarrow 0} [p \, \tilde{\mathbf{C}}(p)] = 
\lim_{p \rightarrow 0} \left[ p \, \mathbf{N} + \mathbf{B} + p \int_{]0,\infty[} (p + r)^{-1} \mathbf{G}(r)\, \mu(\dd r)\right]^{-1}$$
If the limit $\lim_{p \rightarrow 0} \left[ \mathbf{B} + p \int_{]0,\infty[} (p + r)^{-1} \mathbf{G}(r)\, \mu(\dd r)\right]
\equiv \mathbf{B} + \lim_{t\rightarrow \infty} \mathbf{F}_0(t) \equiv \lim_{t\rightarrow \infty} \mathbf{R}(t)$ exists and is invertible,\
then equation~\eqref{eq3} is satisfied.

\mbox{ }\hfill $\Box$

\begin{theorem}
If $\mathbf{C}$ is a $\mathcal{S}_+$-valued Bernstein function, 
and \\
$(\ast\ast)$\; for each non-zero vector $\mathbf{v} \in \mathbb{R}^6$ the function $R_{IJ}(t)\, v_I\, v_J$ is not identically zero, \\
then there is an $\mathcal{S}_+$-valued LICM function $\mathbf{F}$ 
and $\mathbf{N} \in \mathcal{S}_+$ such that
\begin{equation} \label{uu}
\mathbf{R}(t) = u(t)\, \mathbf{N} + \mathbf{F}(t)
\end{equation}
satisfies equation~\eqref{anisorel}. 

The creep modulus has the form
\begin{equation} \label{ha}
\mathbf{C}(t) = \mathbf{A} + t \, \mathbf{B} + \int_0^t \mathbf{Q}(s)\, \dd s,
\end{equation}
where $\mathbf{A}, \mathbf{B} \in \mathcal{S}_+$, $\lim_{t\rightarrow \infty} \mathbf{C}^\prime(t) = \mathbf{B}$  and $\mathbf{Q}$ is a LICM.

If $\mathbf{A} \equiv \mathbf{C}(0)$ has an inverse,  then 
\begin{equation}\label{ww}
\lim_{t\rightarrow 0} \mathbf{R}(t) = \mathbf{C}(0)^{-1}
\end{equation}

If $\mathbf{B} > 0$,  then $\lim_{t\rightarrow \infty} \mathbf{R}(t) = 0$.

If $\mathbf{B} = 0$, the limit of $\mathbf{C}(t)$ at infinity exists and is invertible, then 
\begin{equation} \label{uw}
\lim_{t\rightarrow\infty} \mathbf{R}(t) = \left[ \lim_{t\rightarrow \infty} \mathbf{C}(t)\right]^{-1}
\end{equation}

\end{theorem}

\noindent\textbf{Proof.}\\
Condition $(\ast\ast)$ ensures that the matrix $\tilde{\mathbf{C}}(p)$  is invertible for $p \geq 0$.

We now note that $\mathbf{C}(t) = \int_0^t \mathbf{L}(s)\, \dd s$, where $\mathbf{L}$ is an $\mathcal{S}_+$-valued LICM function.
Consequently $p \, \tilde{\mathbf{C}}(p) = \tilde{\mathbf{L}}(p)$ is an $\mathcal{S}_+$-valued Stieltjes function. Its inverse is an
$\mathcal{S}_+$-valued CBF and it has the form
$$\mathbf{N} + p\, \mathbf{B} + p \int_{]0,\infty[} (p + r)^{-1} \, \mathbf{G}(r)\, \mu(\dd r),$$
with $\mathbf{N}, \mathbf{B} \in \mathcal{S}_+$, a Borel measure $\mu$ satisfying \eqref{x2} and a 
bounded measurable $\mathcal{S}_+$-valued function $\mathbf{G}$ on $]0,\infty[$.
Equation~\eqref{anisorel} is satisfied if 
$\mathbf{R}$ is given by equation~\eqref{uu} with $$\mathbf{F}(t) := \mathbf{B} + 
\int_{]0,\infty[} \e^{-r t} \, \mathbf{G}(r)\, \mu(\dd r).$$ 

Since for each $\mathbf{v} \in \mathbb{R}^6$ the function $\mathbf{v}^\mathsf{T} \, \mathbf{R}(t)\, \mathbf{v}$ 
is CM, it has a non-negative limit at infinity. Hence $\lim_{t\rightarrow \infty} \mathbf{R}(t)$ exists and is positive-semidefinite.

Since $p \, \tilde{\mathbf{C}}(p)$ is a Stieltjes function
$$p \,\tilde{\mathbf{C}}(p) = \mathbf{A} + p^{-1} \, \mathbf{B} + \int_{]0,\infty[} (p +r)^{-1}\, \mathbf{H}(r)\, \nu(\dd r)$$
for some $\mathbf{A}, \mathbf{B} \in \mathcal{S}_+$, a Borel measure $\nu$ satisfying \eqref{x2} 
and a bounded $\mathcal{S}_+$-valued function $\mathbf{H}$. 
Defining a LICM function  $\mathbf{Q}(t) := \int_{]0,\infty[} \e^{-r t}\, \mathbf{H}(r)\, \nu(\dd r)$ 
we obtain equation~\eqref{ha}. 
$\mathbf{C}(t) = \mathbf{A} + t \, \mathbf{B} + \int_0^t \mathbf{Q}(s)\, \dd s$.

Consider
$$\lim_{t\rightarrow \infty} \mathbf{Q}(t) = \lim_{t\rightarrow \infty} \int_{]0,\infty[} \e^{-r t} \, \mathbf{H}(r)\, \nu(\dd r).$$
On account of the inequality $\e^{-x} \leq (1 + x)^{-1}$ for $x > 0$ the integrand in the last integral 
is majorized by $(1 + r)^{-1} \, \nu(\dd r)$ for $t > 1$. On account of \eqref{x2} and the Lebesgue Dominated
Convergence Theorem $\lim_{t\rightarrow \infty} \mathbf{Q}(t) = 0$. 
Consequently the derivative $\mathbf{C}^\prime(t)$ tends to $\mathbf{B}$ at infinity.

If $\mathbf{B} > 0$, then the function $\mathbf{C}(t)$ diverges at infinity.
 
We now investigate the limit $\lim_{t \rightarrow 0} \mathbf{R}(t) = \lim_{p\rightarrow \infty}
\left[\mathbf{A} + p^{-1}\, \mathbf{B} + \tilde{\mathbf{Q}}(p) \right]^{-1}$.
Noting that $\lim_{p\rightarrow \infty} \tilde{\mathbf{Q}}(p) = \lim_{t\rightarrow 0} \int_0^t \mathbf{Q}(s)\, \dd s = 0$ 
(because $\mathbf{Q}$ is locally integrable) we get
$$\lim_{t\rightarrow 0} \mathbf{R}(t) = \mathbf{A}^{-1} \equiv \mathbf{C}(0)^{-1}$$ if $\mathbf{C}(0)$ is invertible..

We now note that
$$\lim_{t \rightarrow \infty} \mathbf{R}(t) = \lim_{p\rightarrow 0} \left\{p \, \left[p \, \mathbf{A} +
\mathbf{B} + p\, \tilde{\mathbf{Q}}(p)\right]^{-1}\right\}.$$ The last term in the square brackets tends to 
$\lim_{t\rightarrow 0} \mathbf{Q}(t) = 0$.

If $\mathbf{B} > 0$, then the expression in the square brackets tends to $\mathbf{B}$, and therefore $\lim_{t \rightarrow \infty} \mathbf{R}(t) = 0$.

If $\mathbf{B} = 0$, then $\lim_{t \rightarrow \infty} \mathbf{R}(t) = 
\lim_{p\rightarrow 0} \left[ \mathbf{A} + \tilde{\mathbf{Q}}(p)\right]^{-1}$.
The limit of the expression in the square brackets is $\mathbf{A} + \int_0^\infty \mathbf{Q}(s)\, \dd s) = \lim_{t\rightarrow \infty} \mathbf{C}(t)$.
If the last limit exists and is invertible, then equation~\eqref{uw} is satisfied.

\mbox{ }\hfill $\Box$

Condition $(\ast)$ ensures that a non-zero strain always causes a non-zero stress.
Condition $(\ast\ast)$ ensures that a non-zero stress always causes a non-zero strain.

\section{Concluding remarks}

We have identified the Laplace transforms of the relaxation modulus and creep function as 
members of appropriate classes of functions. Availability of integral representations for
these classes allows a complete listing of the component terms relevant for the
duality equation. We have thus determined all the components of the relaxation modulus 
and the creep function in a class of viscoelastic models characterized by completely 
monotone relaxation.

The proofs of equation~\eqref{duality} suggest that the relaxation modulus and the 
creep function should be considered as convolution operators. Such an approach allows 
incorporating the identity operator in this class. The identity operator cannot be ignored 
in the context of the duality equation \eqref{duality}. Even though one might set  the 
Newtonian viscosity term $\beta = 0$ in \eqref{f0} or $\mathbf{N} = 0$ in equation~\eqref{uu}, 
a "Newtonian viscosity" term $b\, t$ or $t \, \mathbf{B}$ will pop up in equation~\eqref{ht}.

Given an anisotropic structure of the medium ane might construct the function $\mathbf{F}(t)$
in Theorem~3 by setting $\mathbf{G}(r) = \sum_{J=1}^6 \lambda_J(r)\, \mathbf{S}_J \otimes \mathbf{S}_J$
with $0 \leq \lambda_J(r) \leq(1)$, for $J = 1,\ldots 6$ and the eigenstresses $\mathbf{S}_J$ held
constant. For $\mathbf{F}$ we get the following formula
$$\mathbf{F}(t) = \mathbf{B} + \sum_{J=1}^6 f_J(t) \, \mathbf{S}_J \otimes \mathbf{S}_J$$
with $$f_J(t) = \int_{]0,\infty[} \e^{-r t} \, \lambda_J(r)\, \mu(\dd r)$$ a LICM function.
$f_I(t)$ represents the relaxation of the eigenstress $\Sigma_I$ scaused by a jump
of the eigenstrain $\Epsilon_I$ from 0 to $\Epsilon_I$.

In this case the undrlying anisotropic directional structure is not affected by relaxation and the functions
$f_J$ account for different relaxations of different eigenstresses. The eigenstrains are determined
by the anisotropy class.

\appendix

\section{A few relevant properties of LICM, complete Bernstein and Stieltjes functions.} \label{appscalar}

For details, see \cite{SchillingAl,SerHanJMP}. 

In order to focus on those statements which 
are of use for us we shall consider as definitions some statements that appear as theorems  
in the references cited above.

An infinitely differentiable function $f$ is said to be LICM if it is completely monotone:
$$(-1)^n\, \D^n f(t) \geq 0 \     \text{for}  \      n = 0,1, 2,...$$
and integrable in a neighborhood of 0 \cite{SerHanJMP}.

If $f(t)$ is LICM, then there is a real number $a \geq 0$ and a 
Borel measure $\mu$ on $]0,\infty[$ 
satisfying \eqref{x2} such that
$$f(t) = a + \int_{]0,\infty[} \e^{-r t}\, \mu(\dd r)$$ 
The last equation can also be 
recast in a more familiar form
$$f(t) = \int_{[0,\infty[} \e^{-r t}\, \mu(\dd r)$$
by defining $\mu(\{0\}) = a$. 

The inverse implication is also true.

A Bernstein function is defined as an infinitely differentiable function $g$ on $[0,\infty[$ satisfying the inequalities 
$g(t) \geq 0$ and $(-1)^n \,\D^n g(t) \leq 0$ for $n = 1,2,\ldots$.

If $f$ is LICM, $a, b \geq 0$, then $g(t) = \int_0^t g(s) \, \dd s + a + b\, t$ 
is a Bernstein function. The derivative of a Bernstein function is LICM.

A function $f$ is a Stieltjes function if there are two real numbers $a, b \geq 0$ and
a Borel measure $\mu$ satisfying \eqref{x2} such that 
\begin{equation} \label{app1}
f(p) = a  + \frac{b}{p} + \int_{]0,\infty[}\frac{\mu(\dd r)}{p + r} 
\end{equation}
The integration extends over $0 < r < \infty$. The second term can be incorporated in the integral
by extending the integration to $[0,\infty[$, but we prefer to keep it separate \cite{SchillingAl}, Def.~2.1.

A function $f$ is a CBF if there are two real numbers $a, b \geq 0$ and a Borel measure 
$\nu$ on $]0,\infty[$
satisfying \eqref{x0} 
such that 
\begin{equation}\label{app2}
f(p) = a + b\, p + p \int_{]0,\infty[} \frac{\nu(\dd r)}{p + r}
\end{equation}
\cite{SchillingAl}, Thm~6.2. 

$f(p)$ is a Stieltjes function if and only if $p\, f(p)$ is a CBF.
This statement follows from the integral representations of CBFs and Stieltjes functions above.

The following non-linear relation between the CBFs and the Stieltjes functions is the key to 
the proof above:
A function  $f(p)$ not identically zero is a Stieltjes function if and only if  $1/f(p)$ is 
a CBF not identically 0.
\cite{SchillingAl}, Thm~7.3.

\section{Some relevant properties of matrix-valued functions of the Stieltjes and complete Bernstein class.} \label{appAniso}

For details see \cite{HanAnisoWaves}.

Let $\mathcal{S}_+$ denote the set of non-negative symmetric matrices.

A symmetric matrix-valued
function $\mathbf{A}(t)$ is CM if 
$$ (-1)^n \, \D^n \,\mathbf{A}(t) \geq 0 \hspace{1cm} \text{for} \; n = 0, 1, 2\ldots $$
where $\mathbf{B} \geq 0$ is equivalent to $\mathbf{v}^\mathsf{T}\, \mathbf{M}\, \mathbf{v} \geq 0$ for every 
$\mathbf{v} \in \mathbb{R}^6$.

The function $\mathbf{A}(t)$ is LICM if it is CM and locally integrable. 

If $\mathbf{A}$ is LICM then for every vector $\mathbf{v} \in \mathbb{R}^6$ the function
$\mathbf{v}^\mathsf{T} \, \mathbf{A}\, \mathbf{v}$ is LICM. By Bernstein's theorem
there is a Borel measure $m_\mathbf{v}$ on $[0,\infty[$ such that 
$$\mathbf{v}^\mathsf{T} \, \mathbf{A}(t)\, \mathbf{v} = \int_{[0.\infty[} \e^{-r t}\, m_\mathbf{v}(\dd r)$$
and
\begin{equation}\label{x2a}
\int_{]0.\infty[} (1 + r)^{-1}\, m_\mathbf{v}(\dd r)  < \infty 
\end{equation}
We now note that 
\begin{equation} \label{df}
4 \mathbf{v}^\mathsf{T} \, \mathbf{A}(t) \, \mathbf{w} = \int_{[0,\infty[} \e^{-r t} \, M_{\mathbf{v},\mathbf{w}}(\dd r)
\end{equation}
for every $\mathbf{v}, \mathbf{w} \in \mathbb{R}^6$, where $M_{\mathbf{v},\mathbf{w}}(E) := 
m_{\mathbf{v}+\mathbf{w}}(E) - m_{\mathbf{v}-\mathbf{w}}(E)$ for every Borel $E \subset [0,\infty[$ and 
$\mathbf{v}, \mathbf{w} \in \mathbb{R}^6$. From equation~\eqref{df}, using the uniqueness of the Laplace transform,
follows that $M_{\lambda \, \mathbf{v} + \mathbf{z},\mathbf{w}}(E) = \lambda\, M_{\mathbf{v},\mathbf{w}}(E) + 
M_{\mathbf{z},\mathbf{w}}(E)$
hence there is a matrix $\mathbf{M}(E)$ such that $M_{\mathbf{v},\mathbf{w}}(E) = \mathbf{v}^\mathsf{T}\, \mathbf{M}(E) \,
\mathbf{w}$. This matrix is also symmetric. Since $m_{\mathbf{v}}(E) \geq 0$ and $m_\mathbf{0}(E) = 0$,  
the matrix $\mathbf{E}$ is positive semi-definite.

Furthermore, denoting $\mathbf{M}(E) =: \mathbf{H}$, we note that a square root $\mathbf{H}^{1/2}$ such that
$\mathbf{H} = \mathbf{H}^{1/2} \, \mathbf{H}^{1/2}$ (see, e. g. \cite{HanAnisoWaves} for the definition). 
The square root is a positive semi-definite symmetric matrix, hence
\begin{equation} \label{xy}
\vert \mathbf{v}^\mathsf{T}\,\mathbf{H}\, \mathbf{w} \vert^2 = \left\vert \left(\mathbf{H}^{1/2}\, \mathbf{v}\right)^\mathsf{T}\, 
\left(\mathbf{H}^{1/2}\, \mathbf{w}\right) \right\vert^2 
 \leq [\mathbf{v}^\mathsf{T}\, \mathbf{H}\, \mathbf{v}]\,[\mathbf{w}^\mathsf{T}\, \mathbf{H}\, \mathbf{w}]
\end{equation}
Since $\mathbf{H} \geq 0$, the right-hand side of \eqref{xy} is bounded from above by $\mathrm{trace}(\mathbf{H})^2 \, \vert \mathbf{v} \vert^2 \, \mathbf{w}\vert^2 $. 

Define the Borel measure $\mu(E) := \mathrm{trace}(\mathbf{M}(E))$ for Borel $E \subset [0,\infty[$.
Since $\vert \mathbf{v}^\mathsf{T}\,\mathbf{M}(E)\, \mathbf{w} \vert \leq \mu(E) \, \vert \mathbf{v} \vert
\,  \vert \mathbf{w} \vert$, 
by the Radon-Nikodym theorem there is a bounded measurable function $\mathbf{g}_{\mathbf{v},\mathbf{w}}$ on $[0,\infty[$, defined $\mu$-almost everywhere,
such that $\mathbf{M}(E) = \int_E \mathbf{g}_{\mathbf{v},\mathbf{w}}(r)\, \mu(\dd r)$ for every Borel $E \subset ]0,\infty[$. Repeating an argument used above one can prove that 
$\mathbf{g}_{\mathbf{v},\mathbf{w}}(r) = \mathbf{v}^\mathsf{T} \, \mathbf{G}(r)\, \mathbf{w}$, where $\mathbf{G}(r)$ is an $\mathcal{S}_+$-valued 
function on $[0,\infty[$. We conclude that for every symmetric LICM function $\mathbf{A}$ there is a Borel measure $\mu$ and a
bounded $\mathcal{S}_+$-valued 
function $\mathbf{G}$ on $[0,\infty[$ such that
\begin{equation} \label{mLICM}
\mathbf{A}(t) = \int_{[0,\infty[} \e^{-r t}\, \mathbf{G}(r) \, \mu(\dd r)
\end{equation}

We note that on account of \eqref{x2a} the Borel measure  $\mu$ satisfies equation~\eqref{x2}. 

If $\mu(\{0\}) > 0$, then $\mathbf{G}(0)$ is defined and \eqref{mLICM} can be recast in the form
\begin{equation}\label{mLICM1}
\mathbf{A}(t) = \mathbf{B} + \int_{]0,\infty[} \e^{-r t}\, \mathbf{G}(r) \, \mu(\dd r)
\end{equation}
where $\mathbf{B} := \mu(\{0\})\, \mathbf{G}(0)$ is a positive semi-definite symmetric matrix. If $\mu(\{0\}) = 0$ then $\mathbf{B} = 0$.

Calculation of limits of $\mathbf{A}(t)$ imposes the necessity to split $\mathbf{A}(t)$ into two term and consider 
$\mu$ as a Borel measure on $]0,\infty[$.

The Laplace transform of the $\mathcal{S}_+$-valued LICM $\mathbf{A}(t)$ is given by the equation 
\begin{equation} \label{mlapLICM}
\tilde{\mathbf{A}}(p) = p^{-1}\, \mathbf{B} +  \int_{]0,\infty[} (p + r)^{-1} \, \mathbf{G}(r) \, \mu(\dd r)
\end{equation}

An $\mathcal{S}_+$-valued Bernstein function is an indefinite integral of a $\mathcal{S}_+$-valued LICM function.

We shall now recall some results from Appendix~B of \cite{HanAnisoWaves}.

A matrix-valued Stieltjes function $\mathbf{Y}(p)$ has the following integral representation:
\begin{equation}\label{mStieltjes}
\mathbf{Y}(p) = \mathbf{B} +  p^{-1}\, \mathbf{C} + \int_{]0,\infty[} (p + r)^{-1} \, \mathbf{G}(r) \, \mu(\dd r)
\end{equation}
where $\mathbf{B}, \mathbf{C} \in \mathcal{S}_+$, $\mu$ is a Borel measure on $]0,\infty[$ satisfying \eqref{x2} and 
 $\mathbf{G}(r)$ is an $\mathcal{S}_+$-valued function defined
$\mu$-almost everywhere on $]0,\infty[$. Conversely, any matrix-valued function with the integral representation 
\eqref{mStieltjes} is an $\mathcal{S}_+$-valued Stieltjes function.

An $\mathcal{S}_+$-valued CBF $\mathbf{Z}(p)$ has the following integral representation:
\begin{equation}\label{mCBF}
\mathbf{Z}(p) = \mathbf{B} + p \, \mathbf{C} + p \int_{]0,\infty[} (p + r)^{-1} \, \mathbf{H}(r) \, \nu(\dd r)
\end{equation}
where $\mathbf{B}, \mathbf{C} \in \mathcal{S}_+$, $\nu$ is a Borel measure on $]0,\infty[$ satisfying \eqref{x0} and 
 $\mathbf{H}(r)$ is an $\mathcal{S}_+$-valued function defined
$\nu$-almost everywhere on $]0,\infty[$. Conversely, any matrix-valued function with the integral representation 
\eqref{mCBF} is a $\mathcal{S}_+$-valued CBF.

It follows immediately that the the function $p^{-1} \, \mathbf{Z}(p)$, where $\mathbf{Z}$ is an $\mathcal{S}_+$-valued 
CBF function, is an $\mathcal{S}_+$-valued Stieltjes function. 

According to Lemma~3 op. cit. if $\mathbf{Z}(p)$ is an invertible $\mathcal{S}_+$-valued CBF then $\mathbf{Z}(p)^{-1}$ is an 
$\mathcal{S}_+$-valued Stieltjes function and conversely.

\end{document}